\documentclass{article}
\usepackage{amsfonts}
\usepackage{amsmath}

\newtheorem{theorem}{Theorem}

\newtheorem{definition}[theorem]{Definition}

\newtheorem{proposition}[theorem]{Proposition}

\newenvironment{proof}[1][Proof]{\noindent\textbf{#1.} }{\ \rule{0.5em}{0.5em}}
\input{tcilatex}

\begin{document}

\title{Clipping over dissipation in turbulence models\thanks{%
Department of Mathematics, University of Pittsburgh, Pittsburgh PA 15260,
USA, emails: kkh16@pitt.edu, wjl@pitt.edu, mhs64@pitt.edu. The work of KK
and WL was partially supported by NSF grant DMS 2110379}}
\author{Kiera Kean, William Layton, and Michael Schneier \\
%EndAName
The University of Pittsburgh}
\maketitle

\begin{abstract}
Clipping refers to adding 1 line of code $A\Leftarrow min\{A,B\}$ to force
the variable $A$ to stay below a present bound B. Phenomenological clipping
also occurs in turbulence models to correct for over dissipation caused by
the action of eddy viscosity terms in regions of small scales. Herein we
analyze eddy viscosity model energy dissipation rates with 2
phenomenological clipping strategies. Since the true Reynolds stresses are $%
O(d^{2})$ ($d=$ wall normal distance) in the near wall region, the first is
to force this near wall behavior in the eddy viscosity by $\nu
_{turb}\Leftarrow \min \{\nu _{turb},\frac{\kappa }{T_{ref}}d^{2}\}$ for
some preset $\kappa $ and time scale $T_{ref}$. The second is Escudier's
early proposal to clip the turbulence length scale, reducing too large
values in the interior of the flow. Analyzing respectively shear flow
turbulence and turbulence in a box (i.e., periodic boundary conditions), we
show that both clipping strategies do prevent aggregate over dissipation of
model solutions.
\end{abstract}

\section{Introduction}

\begin{gather*}
\text{\textit{We dedicate this work to Max Gunzburger}. } \\
\text{\textit{He started us on this adventure and inspired us along the way.}%
}
\end{gather*}%
\textit{Clipping} in scientific programming refers to adding 1 line of code
to force a preset upper or lower bound such as $A\Leftarrow \min \{A,B\}$.
As an example, the $\sqrt{k}$\ term in the model (\ref{eq:1EqnModel}) below
is often implemented as $\sqrt{\max \{k,0\}}$ clipping small negative $k$
values. Phenomenologically deduced clipping occurs in turbulence models to
correct for over dissipation caused by the action of eddy viscosity terms in
regions of small velocity scales and is tested in numerical experiments.
Herein we develop a third, analytical support for phenomenological clipping
in turbulence models. We analyze dissipation in clipped URANS (Unsteady
Reynolds Averaged Navier Stokes) models in two cases. The true Reynolds
stresses are $\mathcal{O}(d^{2})$ ($d=\inf \{|x-y|:y\in \partial \Omega \}$
, the wall normal distance) in the near wall region. The first is to force
this $\mathcal{O}(d^{2})$ behavior in the eddy viscosity by $\nu
_{turb}\Leftarrow \min \{\nu _{turb},\frac{\kappa }{T_{ref}}d^{2}\}$ for
some preset and non-dimensional $\kappa $ and time scale $T_{ref}$. The
second is Escudier's clipping of the model's turbulence length scale, (\ref%
{eq:EscudierCap}) below, in the interior. Analyzing respectively shear flow
turbulence and turbulence in a box (i.e., periodic boundary conditions), we
show that \textit{both clipping strategies do prevent aggregate over
dissipation of eddy viscosity model solutions}.

To establish this, we analyze energy dissipation rates for models of
averages of turbulent velocities and pressures. A wide variety of such
models exist but current practice, summarized in Wilcox \cite{Wilcox},
favors eddy viscosity based, URANS models arising from time averaging. For
example, Durbin and Pettersson Reif \cite{DP11} p. 195 write \textit{%
"Virtually all practical engineering computations are done with some
variation of eddy viscosity ...}". Following for example Mohammadi and
Pironneau \cite{MP} and Wilcox \cite{Wilcox} p.37 equation 3.9, the model
velocity $v(x,t)\simeq \overline{u}(x,t)$ approximates the finite time
average\footnote{%
The time average can occur after ensemble averaging plus an ergodic
hypothesis. URANS models are also constructed ad hoc simply by adding $\frac{%
\partial v}{\partial t}$\ to a RANS model.} $\overline{u}$\ of the
Navier-Stokes velocity $u$%
\begin{equation}
\overline{u}(x,y,z,t)=\frac{1}{\tau }\int_{t-\tau }^{t}u(x,y,z,t^{\prime
})dt^{\prime }\text{ and fluctuation }u^{\prime }:=u-\overline{u}.
\end{equation}%
Causality requires the time window, $t-\tau <t^{\prime }<t$, to stretch
backwards as above so present velocities do not depend on future forces. The
associated turbulent kinetic energy is then $\frac{1}{2}\overline{|u-%
\overline{u}|^{2}}$. Averaging the Navier Stokes equations (NSE) yields the
system $\nabla \cdot \overline{u}=0$\ and%
\begin{gather*}
\overline{u}_{t}+\overline{u}\cdot \nabla \overline{u}-\nabla \cdot (2\nu
\nabla ^{s}\overline{u})-\nabla \cdot R(u,u)+\nabla p=\frac{1}{\tau }%
\int_{t-\tau }^{t}f(x,y,z,t^{\prime })dt^{\prime }\text{ } \\
\text{where \ \ }R(u,u)=\overline{u}\otimes \overline{u}-\overline{u\otimes u%
}.
\end{gather*}%
Here $\nu $ is the kinematic viscosity, $p$ is a pressure, $f$ is the body
force, $\nabla ^{s}u$ is the symmetric part of $\nabla u$, $U$ is a global
velocity scale, $L$ is a global length scale\ and the Reynolds number is $%
\mathcal{R}e=LU/\nu $. This equation is not closed. Models replace $R(u,u)$
by terms that only depend on $\overline{u}$. For time window $\tau $
sufficiently large (and $t>\tau $) time dependence disappears from the
equation and steady state RANS models result. For time window small, $\tau $
can be treated as a small parameter in $R(u,u)$\ and models can be derived
by asymptotics. Herein we consider URANS modelling for intermediate $\tau $.

The main URANS model used in practical turbulent flow predictions is of eddy
viscosity type. Its velocity $v(x,t)\simeq \overline{u}(x,t)$ satisfies 
\begin{equation}
v_{t}+v\cdot \nabla v-\nabla \cdot \left( 2[\nu +\nu _{turb}]\nabla
^{s}v\right) +\nabla p=\frac{1}{\tau }\int_{t-\tau }^{t}f(x,y,z,t^{\prime
})dt^{\prime }\text{, }\nabla \cdot v=0  \label{eq:EVmodel}
\end{equation}%
where the eddy or turbulent viscosity $\nu _{turb}(\geq 0)$ must be
specified. A classical turbulent viscosity specification is the $0-$equation
Smagorinsky-Ladyzhenskaya model $\nu _{turb}=(0.1\delta )^{2}|\nabla ^{s}v|$
where $\delta =$ selected length scale, analyzed by Du, Gunzburger and
Turner in \cite{DG91}, \cite{TG88}. The classic $1-$equation model of
Prandtl and Kolmogorov is analyzed in Section 4. $2-$equation models add a
second, phenomenologically derived equation that determines the $1-$equation
turbulence length scale $l$. In all these cases, the total \textit{model
energy dissipation rate per unit volume} is%
\begin{equation}
\varepsilon _{\text{model}}(v):=\frac{1}{|\Omega |}\int_{\Omega }2[\nu +\nu
_{turb}]|\nabla ^{s}v(x,t)|^{2}dx.  \label{eq:epsilonDefn}
\end{equation}%
A common failure mode of eddy viscosity models is over dissipation, either
producing a lower $\mathcal{R}e$\ flow or even driving the solution to a
nonphysical steady state. This occurs due to the action of the turbulent
viscosity term near walls or on interior small scales. We study over
dissipation here through interrogation of the above model energy dissipation
rate. A wide range of boundary conditions occur in practical flow
simulations. Herein we focus on two: shear boundary conditions to study
turbulence generated by near wall flows (Section 3) and $L-$periodic to
study turbulence dynamics away from walls (Section 4).

\textbf{Section 3 studies clipping }$\nu _{turb}$\textbf{\ near wall for
general eddy viscosity models}. The near wall behavior of the true Reynolds
stress is $\nabla \cdot R(u,u)=\mathcal{O}(d^{2})$. Matching this behavior
in the model requires $\nu _{turb}=\mathcal{O}(d^{2})$. Choosing the
(dimensionless) constant $\kappa $ and the reference time $T_{ref}$, this
near wall asymptotics is enforceable through the clipping 
\begin{equation*}
\nu _{turb}\Leftarrow \min \{\nu _{turb},\frac{\kappa }{T_{ref}}d^{2}\}.
\end{equation*}%
The analysis of the effect of this near wall clipping on energy dissipation
is performed in Section 3 for shear flows. Let $\mathcal{\nu }_{eff}$\
denote\ the effective viscosity (so $\frac{\mathcal{\nu }}{\mathcal{\nu }%
_{eff}}\leq 1$) and $\mathcal{R}e_{eff}=LU/\mathcal{\nu }_{eff}$. We prove
in Theorem 3.1 that this forced replication of the near wall asymptotics of
the true Reynolds stresses does preclude model dissipation as long as $%
\kappa \leq \mathcal{O}(\mathcal{R}e_{eff})$. Theorem 3.1 asserts 
\begin{equation*}
\lim \sup_{T\rightarrow \infty }\frac{1}{T}\int_{0}^{T}\varepsilon _{\text{%
model}}(t)dt\leq \left[ \frac{5}{2}+32\frac{\mathcal{\nu }}{\mathcal{\nu }%
_{eff}}+\frac{\kappa }{6}\mathcal{R}e_{eff}^{-1}\left( \frac{T^{\ast }}{%
T_{ref}}\right) \right] \frac{U^{3}}{L}.
\end{equation*}

\textbf{Section 4 studies Escudier's clipping of }$l$\textbf{\ away from
walls when the eddy viscosity is determined through }$1-$\textbf{equation
models}. The standard formulation of $\nu _{turb}$, due to Prandtl and
Kolmogorov, is $\nu _{turb}=\mu l\sqrt{k}$ where $\mu $ (typically $0.2$ to $%
0.6$) is a calibration constant, $l$ is a turbulence length scale and $k$ $%
\simeq $\ $\frac{1}{2}\overline{|u-\overline{u}|^{2}}$ is a model
approximation to turbulent kinetic energy. Escudier observed that the
traditional value $l=0.41d$ is too large in the flow interior. Escudier \cite%
{E66}, \cite{E67} (see also \cite{Wilcox}, p.78 equation 3.108 and Ch. 3,
eqn. (3.99) p.76) proposed clipping its maximum value (with the cap active
away from walls) by%
\begin{equation}
l=\min \{0.41d,0.09\delta \}\text{ where }\delta =\text{ estimate of
transition region width.}  \label{eq:EscudierCap}
\end{equation}%
In Section 4 we analyze the effect of this clipping in the interior of a
turbulent flow via periodic boundary conditions. Theorem 4.1 establishes
that over dissipation is again prevented%
\begin{equation*}
\lim \sup_{T\rightarrow \infty }\frac{1}{T}\int_{0}^{T}\varepsilon _{\text{%
model}}(t)dt\leq \left[ 3+\frac{9}{2}\mathcal{R}e^{-1}+0.03\mu ^{3/2}\left( 
\frac{\delta }{L}\right) ^{2}\right] \frac{U^{3}}{L}.
\end{equation*}

\subsection{Previous work on model development}

Saint-Venant \cite{S43} noted that turbulent mixing increases with "\textit{%
the intensity of the whirling agitation}", \cite{Darrigol}, p.235. Eddy
viscosity models, based on the early work of Saint-Venant's student
Boussinesq \cite{B77}, are based on%
\begin{equation*}
\begin{array}{ccc}
\text{Boussinesq:} &  & 
\begin{array}{c}
\text{\textit{Turbulent fluctuations have }} \\ 
\text{\textit{a dissipative effect on the mean flow.}}%
\end{array}
\\ 
\text{EV hypothesis:} &  & 
\begin{array}{c}
\text{\textit{This dissipation can be modelled by}} \\ 
\text{\textit{\ an eddy viscosity term }}\nabla \cdot (\nu _{turb}\nabla
^{s}\left\langle u\right\rangle )\mathit{.}%
\end{array}%
\end{array}%
\end{equation*}%
Early work in the kinetic theory of gasses suggested the (dimensionally
consistent) relation $\nu _{turb}=\frac{1}{3}lV$ where $V$\ is a velocity
scale and $l$ is an analog to a mean free pass. Prandtl and Kolmogorov noted
that the enhanced mixing of turbulent flows is due to turbulent fluctuations
and concluded that the correct velocity scale should be inferred from the
turbulent kinetic energy $\frac{1}{2}\overline{|u-\overline{u}|^{2}}$. This
reasoning led to the, now universally accepted (and dimensionally
consistent), Kolmogorov-Prandtl relation $\nu _{turb}=\mu l\sqrt{k}$ where 
\begin{equation*}
l:\text{turbulence length scale and }k\text{ }:\text{model approximation to }%
\frac{1}{2}\overline{|u-\overline{u}|^{2}}\text{.}
\end{equation*}%
Pope \cite{Pope} calculates $\mu =0.55$ from the law of the wall. Davidson 
\cite{D15} p. 114, eqn. (4.11a) calculates $\mu \simeq 0.33$\ in 2d and $\mu
\simeq 0.27$\ in $3d$ using a kinetic theory analogy. Prandtl and
Kolmogorov, e.g., \cite{P45}, \cite{CL} p.99, Section 4.4, \cite{D15}, \cite%
{MP} p.60, Section 5.3 or \cite{Pope} p.369, Section 10.3, independently
derived the following equation for the approximation to the turbulent
kinetic energy%
\begin{equation}
k_{t}+v\cdot \nabla k-\nabla \cdot \left( \left[ \nu +\nu _{turb}\right]
\nabla k\right) +\frac{1}{l}k\sqrt{k}=2\nu _{turb}|\nabla ^{s}v|^{2}.
\label{eq:kEquation}
\end{equation}%
The turbulence length scale $l$ was postulated by Taylor in 1915 \cite{T15}
and described by Prandtl \cite{P26} as%
\begin{gather*}
\text{"\textit{... the diameter of the masses of fluid}} \\
\text{\textit{\ moving as a whole in each individual case"}}\mathit{..}
\end{gather*}%
The idea behind $l=0.41d$\ (among many variants \cite{KKG17}) was that near
walls, the \textit{diameter} of a coherent mass of fluid was constrained by
the near wall distance. Away from walls, $l=0.41d$ is too large and Escudier
proposed the cap (\ref{eq:EscudierCap}). Prandtl \cite{P26} in 1926 also
mentioned a second, kinematic possibility\ 
\begin{gather*}
\text{ "...\textit{or again, as the distance traversed by a mass of this type%
}} \\
\text{\textit{\ before it becomes blended in with neighboring masses..."}\ }
\end{gather*}%
This second possibility is a kinematic description of the \textit{distance a
fluctuating eddy travels in one time unit} and motivated the choice $l=\sqrt{%
2}k^{1/2}\tau $ in \cite{JL14b}, \cite{TC04}, \cite{LM18}. Kolmogorov
inferred $l$ from a second equation, beginning the development of $2-$%
equation models. There are many other proposed mixing lengths; the paper 
\cite{KKG17} studies 9 and describes more.

\subsection{Previous work on energy dissipation rates}

The energy dissipation rate is a fundamental statistic of turbulence, e.g., 
\cite{Pope}, \cite{V15}. The balance of energy dissipation with energy
input, $\varepsilon \simeq U^{3}/L$ , is observed in physical experiments 
\cite{Frisch}, \cite{V15}. In 1992 Constantin and Doering \cite{CD92}
established a direct link between phenomenology and NSE predicted energy
dissipation. This work builds on \cite{B78}, \cite{H72} (and others) and has
developed in many important directions subsequently e.g., \cite{DF02}, \cite%
{V15}, \cite{Wang97}, \cite{W00}.

Extending this work to turbulence models requires existence of weak
solutions and a standard energy inequality. An existence theorem for weak
solutions to a general eddy viscosity model is proven in \cite{LL02} in
which the uniform bound on $\nu _{turb}$ induced by clipping automatically
enforces two of the three needed assumptions. The third depends on the
specific dependence of $\nu _{turb}$\ on $v$. The current state of existence
theory is treated comprehensively in \cite{CL}. For some simple turbulence
models, existence is known and \'{a} priori analysis has shown that $%
avg(\varepsilon )\leq \mathcal{O}(U^{3}/L)$, where the hidden constant does
not blow up as $\mathcal{R}e\rightarrow \infty $, e.g., \cite{D16}, \cite%
{L02}, \cite{L07}, \cite{L16}, \cite{LRS10}, \cite{P17}, \cite{P19}, \cite%
{P19b}. For the 1-equation model with length scale $l=\sqrt{2}k^{1/2}\tau $
existence is plausible but still an open problem. Assuming existence and an
energy inequality, this model has been proven in \cite{LM18} not to over
dissipate due to small scales generated by the nonlinearity. In the
Smagorinsky model, Pakzad \cite{P17} has proven that wall damping functions,
a clipping alternative, prevent\ over dissipation.

\section{Notation and preliminaries}

We assume that weak solutions of the systems studied exist and satisfy
standard energy inequalities. In many cases this plausible assumption has
not yet been proven, see \cite{CL} for current knowledge. The $L^{2}(\Omega
) $ norm and the inner product are $\Vert \cdot \Vert $ and $(\cdot ,\cdot )$%
. Likewise, the $L^{p}(\Omega )$ norms is $\Vert \cdot \Vert _{L^{p}}$. $C$
represents a generic positive constant independent of $\nu ,\mathcal{R}e$,
other model parameters and the flow scales $U,L$\ defined below. In all
cases the turbulent viscosity $\nu _{turb}=\nu _{turb}(x,y,z,t)$ and will be
abbreviated by writing $\nu _{turb}$.

\begin{definition}
The finite and long time averages of a function $\phi (t)$\ are%
\begin{equation*}
\left\langle \phi \right\rangle _{T}=\frac{1}{T}\int_{0}^{T}\phi (t)dt\text{
and }\left\langle \phi \right\rangle =\lim \sup_{T\rightarrow \infty
}\left\langle \phi \right\rangle _{T}.
\end{equation*}
\end{definition}

These satisfy $\left\langle \left\langle \phi \right\rangle \right\rangle
=\left\langle \phi \right\rangle $ and 
\begin{equation}
\left\langle \phi \psi \right\rangle _{T}\leq \left\langle |\phi
|^{2}\right\rangle _{T}^{1/2}\left\langle |\psi |^{2}\right\rangle _{T}^{1/2}%
\text{ and }\left\langle \phi \psi \right\rangle \leq \left\langle |\phi
|^{2}\right\rangle ^{1/2}\left\langle |\psi |^{2}\right\rangle ^{1/2}.
\end{equation}

\section{Clipping $\protect\nu _{turb}$ in the turbulent boundary layer}

Over dissipation is often due to incorrect values of $\nu _{turb}$ in
regions of small scales, i.e. where $\nabla ^{s}v$ is large. These small
scales are generated in the boundary layer and in the interior by breakdown
of large scales through the nonlinearity. This section considers those
generated predominantly in the turbulent boundary layer, studied via shear
boundary conditions. Matching the near wall behavior $R(u,u)=\mathcal{O}%
(d^{2})$ in the model's eddy viscosity term requires $\nu _{turb}=\mathcal{O}%
(d^{2})$, enforced through the clipping 
\begin{gather}
\nu _{turb}\Leftarrow \min \{\nu _{turb},\frac{\kappa }{T_{ref}}d^{2}\}\text{
}  \label{eq:NearWallClipping} \\
\text{so that }0\leq \nu _{turb}\leq \frac{\kappa }{T_{ref}}d^{2}.  \notag
\end{gather}

We study the effect of (\ref{eq:NearWallClipping}) via shear flows. Shear
flows can develop several ways. Inflow boundary conditions can emulate a jet
of water entering a vessel. A body force $f(\cdot )$ can be specified to be
non-zero large and tangential at a fixed wall. The simplest (chosen herein)
is a moving wall modelled by a boundary condition $v=g$ on the boundary
where $g\cdot n=0$. This setting includes flows between rotating cylinders.
Select the flow domain $\Omega =(0,L)^{3}$, $L-$periodic boundary conditions
in $x,y$, a fixed-wall no-slip condition at $z=0$ and a wall at $z=L$ moving
with velocity $(U,0,0)$:%
\begin{equation}
\begin{array}{cc}
Boundary & Conditions: \\ 
\text{moving top lid:} & v(x,y,L,t)=(U,0,0) \\ 
\text{fixed bottom wall:} & v(x,y,0,t)=0 \\ 
\text{periodic side walls:} & 
\begin{array}{c}
v(x+L,y,z,t)=v(x,y,z,t), \\ 
v(x,y+L,z,t)=v(x,y,z,t)%
\end{array}%
\end{array}
\label{eq:Shear}
\end{equation}%
Herein, we assume that a weak solution of the model (\ref{eq:EVmodel}) with
shear boundary conditions (\ref{eq:Shear}) exists and satisfies the usual
energy inequality. Specifically, for any divergence free function $\phi (x)$
with $\phi ,\nabla \phi \in L^{2}(\Omega )$ and satisfying the shear
boundary conditions (\ref{eq:Shear}),%
\begin{gather}
\frac{1}{2}\frac{d}{dt}||v||^{2}+\int_{\Omega }2[\nu +\,\nu
_{turb}]|\,\nabla ^{s}{v}|^{2}dx\leq  \label{eq:ShearEnergyIneq} \\
(v_{t},\phi )+\int_{\Omega }2[\nu +\,\nu _{turb}]\nabla ^{s}{v}:\nabla
^{s}\phi dx+(v\cdot \nabla v,\phi ).  \notag
\end{gather}

To formulate our first main result we recall the definition of the \textbf{%
effective viscosity} $\nu _{eff}$ ($\geq \nu $), well defined due to
Proposition 3.3, and a few related quantities.

\begin{definition}
The \textit{effective viscosity }$\nu _{eff}$ is 
\begin{equation*}
\nu _{eff}:=\frac{\left\langle \frac{1}{|\Omega |}\int_{\Omega }[2\nu +2\nu
_{turb}]|\nabla ^{s}v|^{2}dx\right\rangle }{\left\langle \frac{1}{|\Omega |}%
\int_{\Omega }|\nabla ^{s}v|^{2}dx\right\rangle }.
\end{equation*}%
The large scale turnover time is $T^{\ast }=L/U$. The \textit{Reynolds number%
} and \textit{effective Reynolds number} are $\mathcal{R}e=U\,L/\nu $ and\ $%
\mathcal{R}e_{eff}=U\,L/\nu _{eff}.$ Let $\beta =\frac{1}{8}\mathcal{R}%
e_{eff}^{-1}$ and denote the region $\mathcal{S}_{\beta }$ by 
\begin{equation*}
\mathcal{S}_{\beta }=\left\{ (x,y,z):0\leq x\leq L,0\leq y\leq L,(1-\beta
)L<z<L\right\} .
\end{equation*}
\end{definition}

Theorem 3.2 asserts that $\nu _{eff}$ matching the near wall asymptotics of $%
R(u,u)$ is enough to ensure that the model does not over dissipate.

\begin{theorem}
Assume $0\leq \nu _{turb}(x,t)\leq \frac{\kappa }{T_{ref}}d^{2}$. Then, any
weak solution of the eddy viscosity model (\ref{eq:EVmodel}) satisfying the
energy inequality (\ref{eq:ShearEnergyIneq}) has its model energy
dissipation bounded as 
\begin{equation*}
\left\langle \varepsilon _{\text{model}}\right\rangle \leq \left[ \frac{5}{2}%
+32\frac{\mathcal{\nu }}{\mathcal{\nu }_{eff}}+\frac{\kappa }{6}\mathcal{R}%
e_{eff}^{-1}\left( \frac{T^{\ast }}{T_{ref}}\right) \right] \frac{U^{3}}{L}.
\end{equation*}
\end{theorem}

To begin the proof, we recall that uniform bounds follow from (\ref%
{eq:ShearEnergyIneq}) by a known argument.

\begin{proposition}[Uniform Bounds]
Consider the model (\ref{eq:EVmodel}) with shear boundary conditions (\ref%
{eq:Shear}). Assume that there is a $\kappa \geq 0$ such that%
\begin{equation*}
0\leq \nu _{turb}(x,t)\leq \frac{\kappa }{T_{ref}}d^{2}.
\end{equation*}%
Then, for a weak solution satisfying (\ref{eq:ShearEnergyIneq})\ the
following are uniformly bounded in $T$ 
\begin{equation*}
||v(T)||^{2},\int_{\Omega }\,\nu _{turb}(\cdot ,T)dx,\left\langle
\int_{\Omega }|\,\nabla ^{s}{v}|^{2}dx\right\rangle _{T}\text{\ }%
,\left\langle \int_{\Omega }[2\nu +\,2\nu _{turb}]|\,\nabla ^{s}{v}%
|^{2}dx\right\rangle _{T}.
\end{equation*}
\end{proposition}

\begin{proof}
Due to the clipping imposed we have $0<\nu \leq 2\nu +\,2\nu _{turb}\leq
C<\infty $. Since $2\nu +\,2\nu _{turb}$ is positive and uniformly bounded
the above uniform bounds follow from differential inequalities exactly as in
the NSE case and along the lines of the analogous proof in \cite{KLS21}.
\end{proof}

\begin{proof}[Proof of Theorem 3.2]
Following Doering and Constantin \cite{DC92}, choose $\phi (z)=[\widetilde{%
\phi }(z),0,0]^{T}$\ where 
\begin{equation*}
\widetilde{\phi }(z)=\left\{ 
\begin{array}{cc}
0, & z\in \lbrack 0,L-\beta \,L] \\ 
\frac{U}{\beta \,L}(z-(L-\beta \,L)), & z\in \lbrack L-\beta \,L,L]%
\end{array}%
\right. \beta =\frac{1}{8}\mathcal{R}e_{eff}^{-1}.
\end{equation*}%
This function $\phi (z)$ is piecewise linear, continuous, divergence free
and satisfies the boundary conditions. The following are easily calculated
values 
\begin{equation*}
\begin{array}{cc}
||\,\phi \,||_{L^{\infty }(\Omega )}=U, & ||\,\nabla \phi \,||_{L^{\infty
}(\Omega )}=\frac{U}{\beta \,L},\text{ } \\ 
||\,\phi \,||^{2}=\frac{1}{3}\,U^{2}\,\beta \,L^{3}, & \text{ }||\,\nabla
\,\phi \,||^{2}=\frac{U^{2}\,L}{\beta }.%
\end{array}%
\end{equation*}%
With this choice of $\phi $, time averaging the energy inequality (\ref%
{eq:ShearEnergyIneq}) over $[0,T]$ and normalizing by $|\Omega |=L^{3}$
gives 
\begin{gather}
\frac{1}{2TL^{3}}||v(T)||^{2}+\left\langle \frac{1}{L^{3}}\int_{\Omega
}[2\nu +2\,\nu _{turb}]|\,\nabla ^{s}{v}|^{2}dx\right\rangle _{T} \\
\leq \frac{1}{2TL^{3}}||v(0)||^{2}+\frac{1}{TL^{3}}(v(T)-v(0),\phi
)+\left\langle \frac{1}{L^{3}}(v\cdot \nabla v,\phi )\right\rangle _{T} 
\notag \\
+\left\langle \frac{1}{L^{3}}\int_{\Omega }[2\nu +\,2\nu _{turb}]\nabla ^{s}{%
v}:\nabla ^{s}\phi dx\right\rangle _{T}.  \notag
\end{gather}%
Due to the above \'{a} priori bounds the averaged energy inequality can be
written as%
\begin{equation}
\left\langle \varepsilon \right\rangle _{T}\leq \mathcal{O}(\frac{1}{T}%
)+\left\langle \frac{1}{L^{3}}(v\cdot \nabla v,\phi )\right\rangle
_{T}+\left\langle \frac{1}{L^{3}}\int_{\Omega }[2\nu +2\,\nu _{turb}]\nabla
^{s}{v}:\nabla ^{s}\phi dx\right\rangle _{T}.
\end{equation}%
The main issue is thus the third term, $\int \,2\nu _{turb}\nabla ^{s}{v}%
:\nabla ^{s}\phi dx$. Before treating that we recall the analysis of Doering
and Constantine \cite{DC92} and Wang \cite{Wang97} for the two terms shared
by the NSE, $(v\cdot \nabla v,\phi )$\ and $\int 2\nu \nabla ^{s}{v}:\nabla
^{s}\phi dx$. For the nonlinear term $\left\langle \frac{1}{L^{3}}(v\cdot
\nabla v,\phi )\right\rangle _{T}=:NLT$, we have%
\begin{gather*}
NLT=\left\langle \frac{1}{L^{3}}(v\cdot \nabla v,\phi )\right\rangle
_{T}=\left\langle \frac{1}{L^{3}}([v-\phi ]\cdot \nabla v,\phi
)\right\rangle _{T}+\left\langle \frac{1}{L^{3}}(\phi \cdot \nabla v,\phi
)\right\rangle _{T} \\
\leq \left\langle \frac{1}{L^{3}}\int_{\mathcal{S}_{\beta }}|v-\phi ||\nabla
v||\phi |+|\phi |^{2}|\nabla v|dx\right\rangle _{T} \\
\leq \frac{1}{L^{3}}\left\langle 
\begin{array}{c}
\left\Vert \frac{v-\phi }{L-z}\right\Vert _{L^{2}(\mathcal{S}_{\beta
})}||\nabla v||_{L^{2}(\mathcal{S}_{\beta })}||(L-z)\phi ||_{L^{\infty }(%
\mathcal{S}_{\beta })}+ \\ 
+||\phi ||_{L^{\infty }(\mathcal{S}_{\beta })}^{2}||\nabla v||_{L^{1}(%
\mathcal{S}_{\beta })}%
\end{array}%
\right\rangle _{T}.
\end{gather*}%
On the RHS, $||\phi ||_{L^{\infty }(\mathcal{S}_{\beta })}^{2}=U^{2}$and $%
||(L-z)\phi ||_{L^{\infty }(\mathcal{S}_{\beta })}=\frac{1}{4}\beta LU.$
Since $v-\phi $\ vanishes on $\partial \mathcal{S}_{\beta }$, Hardy's
inequality, the triangle inequality and a calculation imply 
\begin{eqnarray*}
\left\Vert \frac{v-\phi }{L-z}\right\Vert _{L^{2}(\mathcal{S}_{\beta })}
&\leq &2\left\Vert \nabla (v-\phi )\right\Vert _{L^{2}(\mathcal{S}_{\beta })}
\\
&\leq &2\left\Vert \nabla v\right\Vert _{L^{2}(\mathcal{S}_{\beta
})}+2\left\Vert \nabla \phi \right\Vert _{L^{2}(\mathcal{S}_{\beta })} \\
&\leq &2\left\Vert \nabla v\right\Vert _{L^{2}(\mathcal{S}_{\beta })}+2U%
\sqrt{\frac{L}{\beta }}.
\end{eqnarray*}%
Thus we have the estimate%
\begin{eqnarray}
NLT &\leq &\frac{\beta LU}{4}\frac{1}{L^{3}}\left\langle 2||\nabla
v||_{L^{2}(\mathcal{S}_{\beta })}^{2}+2U\sqrt{\frac{L}{\beta }}||v||_{L^{2}(%
\mathcal{S}_{\beta })}\right\rangle _{T}  \label{eq:NLTest} \\
&&+\frac{U^{2}}{L^{3}}\left\langle ||\nabla v||_{L^{1}(\mathcal{S}_{\beta
})}\right\rangle _{T}.  \notag
\end{eqnarray}%
For the last term on the RHS, H\"{o}lders inequality in space then in time
implies%
\begin{eqnarray*}
\frac{U^{2}}{L^{3}}\left\langle ||\nabla v||_{L^{1}(\mathcal{S}_{\beta
})}\right\rangle _{T} &=&\frac{U^{2}}{L^{3}}\left\langle \int_{\mathcal{S}%
_{\beta }}|\nabla v|\cdot 1dx\right\rangle _{T} \\
&\leq &\frac{U^{2}}{L^{3}}\left\langle \sqrt{\int_{\mathcal{S}_{\beta
}}|\nabla v|^{2}dx}\sqrt{\beta L^{3}}\right\rangle _{T} \\
&\leq &\frac{U^{2}\sqrt{\beta }}{L^{3/2}}\left\langle \sqrt{\int_{\mathcal{S}%
_{\beta }}|\nabla v|^{2}dx}\right\rangle _{T} \\
&\leq &\frac{U^{2}\sqrt{\beta }}{L^{3/2}}\left\langle \int_{\mathcal{S}%
_{\beta }}|\nabla v|^{2}dx\right\rangle _{T}^{1/2}.
\end{eqnarray*}%
Increase the integral from $\mathcal{S}_{\beta }$ to $\Omega $, use (as $%
\nabla \cdot v=0$) $||\nabla v||^{2}=2||\nabla ^{s}v||^{2}$ and $\beta =%
\frac{1}{8}\mathcal{R}e_{eff}^{-1}.$ Rearranging and using the
arithmetic-geometric inequality gives%
\begin{gather*}
\frac{U^{2}}{L^{3}}\left\langle ||\nabla v||_{L^{1}(\mathcal{S}_{\beta
})}\right\rangle _{T}\leq U^{2}\sqrt{\beta }\left\langle \frac{1}{L^{3}}%
\int_{\Omega }2|\nabla ^{s}v|^{2}dx\right\rangle _{T}^{1/2} \\
\leq U^{2}\sqrt{\frac{2}{8}\frac{1}{LU}}\left\langle \frac{1}{L^{3}}%
\int_{\Omega }\nu _{eff}|\nabla ^{s}v|^{2}dx\right\rangle _{T}^{1/2} \\
\leq \left( \frac{U^{3}}{L}\right) ^{1/2}\frac{1}{2}\left\langle \frac{1}{%
L^{3}}\int_{\Omega }\nu _{eff}|\nabla ^{s}v|^{2}dx\right\rangle _{T}^{1/2} \\
\leq \frac{1}{2}\frac{U^{3}}{L}+\frac{1}{8}\left\langle \frac{1}{L^{3}}%
\int_{\Omega }\nu _{eff}|\nabla ^{s}v|^{2}dx\right\rangle _{T}.
\end{gather*}%
Similar manipulations yield%
\begin{gather*}
\frac{1}{4}\beta LU\frac{1}{L^{3}}\left\langle 2U\sqrt{\frac{L}{\beta }}%
||v||_{L^{2}(\mathcal{S}_{\beta })}\right\rangle _{T}\leq \frac{1}{2}\beta
LU\left\langle \frac{1}{L^{3}}||\nabla v||_{L^{2}(\mathcal{S}_{\beta
})}^{2}\right\rangle _{T}+\frac{1}{8}\frac{U^{3}}{L} \\
\leq \frac{1}{8}\left\langle \frac{1}{L^{3}}\nu _{eff}||\nabla
^{s}v||_{L^{2}(\mathcal{S}_{\beta })}^{2}\right\rangle _{T}+\frac{1}{8}\frac{%
U^{3}}{L}.
\end{gather*}%
Using the last two estimates in the $NLT$ upper bound (\ref{eq:NLTest}), we
obtain%
\begin{equation*}
NLT\leq 2\beta \frac{LU}{\nu _{eff}}\left\langle \frac{1}{L^{3}}\nu
_{eff}||\nabla ^{s}v||_{L^{2}(\mathcal{S}_{\beta })}^{2}\right\rangle _{T}+%
\frac{5}{8}\frac{U^{3}}{L}.
\end{equation*}%
Thus,%
\begin{eqnarray*}
\left\langle \varepsilon \right\rangle _{T} &\leq &\mathcal{O}(\frac{1}{T})+%
\frac{1}{4}\left\langle \frac{1}{L^{3}}\nu _{eff}||\nabla
^{s}v||_{L^{2}(\Omega )}^{2}\right\rangle _{T}+\frac{5}{8}\frac{U^{3}}{L} \\
&&+\left\langle \frac{1}{L^{3}}\int_{\Omega }2[\nu +\,\nu _{turb}]\nabla ^{s}%
{v}:\nabla ^{s}\phi dx\right\rangle _{T}.
\end{eqnarray*}%
Consider now the last term on the RHS. Since $\phi $ is zero off $\mathcal{S}%
_{\beta }$, 
\begin{gather*}
\left\langle \frac{1}{L^{3}}\int_{\Omega }2[\nu +\,\nu _{turb}]\nabla ^{s}{v}%
:\nabla ^{s}\phi dx\right\rangle _{T}=\left\langle \frac{1}{L^{3}}\int_{%
\mathcal{S}_{\beta }}2[\nu +\,\nu _{turb}]\nabla ^{s}{v}:\nabla ^{s}\phi
dx\right\rangle _{T} \\
\leq \frac{1}{2}\left\langle \varepsilon \right\rangle _{T}+\frac{1}{2}%
\left\langle \frac{1}{L^{3}}\int_{\mathcal{S}_{\beta }}2[\nu +\,\nu
_{turb}]\left( \frac{U}{\beta L}\right) ^{2}dx\right\rangle _{T} \\
\leq \frac{1}{2}\left\langle \varepsilon \right\rangle _{T}+\frac{1}{2}%
\left( \frac{U}{\beta L}\right) ^{2}\beta \left\langle \frac{1}{\beta L^{3}}%
\int_{\mathcal{S}_{\beta }}2[\nu +\,\nu _{turb}]dx\right\rangle _{T}.
\end{gather*}%
Thus, as $\beta =\frac{1}{8}\mathcal{R}e_{eff}^{-1}$\ implies $2\beta 
\mathcal{R}e_{eff}=1/4$, 
\begin{eqnarray*}
\frac{1}{2}\left\langle \varepsilon \right\rangle _{T} &\leq &\mathcal{O}(%
\frac{1}{T})+\frac{1}{4}\left\langle \frac{1}{L^{3}}\nu _{eff}||\nabla
^{s}v||_{L^{2}(\Omega )}^{2}\right\rangle _{T} \\
&&+\frac{5}{8}\frac{U^{3}}{L}+\frac{\beta }{2}\left( \frac{U}{\beta L}%
\right) ^{2}\left\langle \frac{1}{\beta L^{3}}\int_{\mathcal{S}_{\beta
}}2\nu +\,2\nu _{turb}dx\right\rangle _{T}.
\end{eqnarray*}%
As a subsequence $T_{j}\rightarrow \infty $%
\begin{equation*}
\left\langle \frac{1}{L^{3}}\nu _{eff}||\nabla ^{s}v||_{L^{2}(\Omega
)}^{2}\right\rangle _{T}\rightarrow \left\langle \varepsilon \right\rangle .
\end{equation*}%
For the other term we calculate%
\begin{eqnarray*}
\left\langle \frac{1}{\beta L^{3}}\int_{\mathcal{S}_{\beta }}\,2\nu
_{turb}dx\right\rangle _{T} &\leq &\left\langle \frac{1}{\beta L^{3}}\int_{%
\mathcal{S}_{\beta }}\,\frac{2\kappa }{T_{ref}}d^{2}dx\right\rangle _{T}=%
\frac{2\kappa }{3T_{ref}}\beta ^{2}L^{2} \\
&=&\frac{2\kappa }{3T_{ref}}L^{2}\left( \frac{1}{8}\mathcal{R}%
e_{eff}^{-1}\right) ^{2}=\frac{2\kappa L^{2}}{192T_{ref}}\mathcal{R}%
e_{eff}^{-2} \\
&&
\end{eqnarray*}%
Thus,%
\begin{equation*}
\frac{1}{2}\left\langle \varepsilon \right\rangle \leq \frac{1}{4}%
\left\langle \varepsilon \right\rangle +\frac{5}{8}\frac{U^{3}}{L}+\frac{1}{2%
}\frac{U^{2}}{\beta L^{2}}\left[ 2\nu +\frac{2\kappa L^{2}}{192T_{ref}}%
\mathcal{R}e_{eff}^{-2}\right] .
\end{equation*}%
Next, insert $\beta =\frac{1}{8}\mathcal{R}e_{eff}^{-1}=\frac{1}{8}\frac{%
\mathcal{\nu }_{eff}}{LU}$ and rewrite $1/T_{ref}=(T^{\ast }/T_{ref})\cdot
(1/T^{\ast })=(T^{\ast }/T_{ref})\cdot (U/L).$ This, after simplification,
completes the proof%
\begin{equation*}
\left\langle \varepsilon \right\rangle \leq \left[ \frac{5}{2}+32\frac{%
\mathcal{\nu }}{\mathcal{\nu }_{eff}}+\frac{\kappa }{6}\mathcal{R}%
e_{eff}^{-1}\left( \frac{T^{\ast }}{T_{ref}}\right) \right] \frac{U^{3}}{L}.
\end{equation*}
\end{proof}

\section{Escudier's clipping of $l$ away from walls}

We analyze in this section the effect, away from walls, of the clipping
developed by Escudier \cite{E66}, \cite{E67}, and still current practice 
\cite{Wilcox}, p.78 equation 3.108 and Ch. 3, eqn. (3.99) p.76,%
\begin{equation*}
l\Leftarrow \min \{l,0.09\delta \}\text{ where }\delta =\text{estimate of
transition layer width}
\end{equation*}%
which implies $0\leq l\leq 0.09\delta $. We prove below that for problems
without boundary layers (studied through periodic boundary conditions) this
cap ensures that energy dissipation rates scale correctly. Thus such caps
are an effective tool for precluding aggregate over dissipation due to the
action of eddy viscosity in regions of interior small scales. Escudier's
proposal (and our analysis) is for the $1-$equation model of Prandtl \cite%
{P45} and Kolmogorov, see also \cite{CL} p.99, Section 4.4, \cite{D15}, \cite%
{MP} p.60, Section 5.3 or \cite{Pope} p.369, Section 10.3, given by%
\begin{gather}
\nabla \cdot v=0,  \notag \\
v_{t}+v\cdot \nabla v-\nabla \cdot \left( \lbrack 2\nu +2\nu _{turb}]\nabla
^{s}v\right) +\nabla p=f(x,y,z),  \label{eq:1EqnModel} \\
k_{t}+v\cdot \nabla k-\nabla \cdot \left( \left[ \nu +\nu _{turb}\right]
\nabla k\right) +\frac{1}{l}k\sqrt{k}=2\nu _{turb}|\nabla ^{s}v|^{2}  \notag
\\
\text{where }\nu _{turb}=\mu l\sqrt{k},\text{ with }\mu \simeq 0.55,\text{
and }0\leq l\leq 0.09\delta .  \notag
\end{gather}%
The flow domain is $\Omega =(0,L_{\Omega })^{3}.$ Since the effect of the
clipped value is in the flow interior, we impose periodic boundary
conditions on $\phi =k(x,t)$ and periodic with zero mean boundary conditions
on $\phi =v,p,v_{0},f$:%
\begin{equation}
\text{\textit{Periodic}: }\phi (x+L_{\Omega }e_{j},t)=\phi (x,t)\text{ and 
\textit{Zero mean}:}\int_{\Omega }\phi dx=0\,.  \label{eq:PeriodicZeroMean}
\end{equation}%
The global length scale $L$ must reflects the scales where the body force is
inputting energy. Define the global velocity scale $U$, the body force scale 
$F$ and large length scale $L$ by%
\begin{equation}
\left. 
\begin{array}{c}
F=\left( \frac{1}{|\Omega |}\int_{\Omega }|f(x)|^{2}dx\right) ^{1/2}\text{, }
\\ 
L=\min \left[ L_{\Omega },\frac{F}{\sup_{x\in \Omega }|\nabla ^{s}f(x)|},%
\frac{F}{\left( \frac{1}{|\Omega |}\int_{\Omega }|\nabla
^{s}f(x)|^{2}dx\right) ^{1/2}}\right] \\ 
U=\left\langle \frac{1}{|\Omega |}\int_{\Omega }|v(x,t)|^{2}dx\right\rangle
^{1/2}.%
\end{array}%
\right\}  \label{eq:FLUdefinition}
\end{equation}%
$L$ has units of length and satisfies%
\begin{equation}
||\nabla ^{s}f||_{\infty }\leq \frac{F}{L}\text{ and }\frac{1}{|\Omega |}%
||\nabla ^{s}f||^{2}\leq \frac{F^{2}}{L^{2}}\text{ }.  \label{eq:PropertiesL}
\end{equation}%
The standard energy inequality and equality for this system are 
\begin{equation}
\left. 
\begin{array}{c}
\frac{d}{dt}\frac{1}{|\Omega |}\frac{1}{2}||v||^{2}+\frac{1}{|\Omega |}%
\int_{\Omega }[2\nu +2\nu _{turb}]|\nabla ^{s}v(x,t)|^{2}dx\leq \frac{1}{%
|\Omega |}(f,v), \\ 
\frac{d}{dt}\int_{\Omega }kdx+\int_{\Omega }\frac{1}{l}k\sqrt{k}%
dx=\int_{\Omega }2\nu _{turb}|\nabla ^{s}v|^{2}dx.%
\end{array}%
\right\}  \label{eq:EnergyIneq1EqnModel}
\end{equation}%
Since $l\leq 0.09\delta $ the following two inequalities hold%
\begin{gather}
\varepsilon _{\text{model}}(t)=\frac{1}{|\Omega |}\int_{\Omega }2[\nu +\nu
_{turb}]|\nabla ^{s}v|^{2}dx\leq \frac{1}{|\Omega |}\int_{\Omega }2\left[
\nu +\mu 0.09\delta \sqrt{k}\right] |\nabla ^{s}v|^{2}dx,  \notag \\
(0.09\delta )^{-1}\left\langle \int_{\Omega }k^{3/2}dx\right\rangle \leq
\left\langle \int_{\Omega }\frac{1}{l}k\sqrt{k}dx\right\rangle =\left\langle
\int_{\Omega }2\nu _{turb}|\nabla ^{s}v|^{2}dx\right\rangle .
\label{eq:AnIneq}
\end{gather}

\begin{theorem}
Consider the $1-$equation model under periodic with zero mean boundary
conditions with $0\leq l\leq 0.09\delta $. The time averaged energy
dissipation rate of any weak solution satisfying the energy inequality (\ref%
{eq:EnergyIneq1EqnModel}) is bounded by%
\begin{equation*}
\left\langle \varepsilon _{\text{model}}\right\rangle \leq \left[ 3+\frac{9}{%
2}\mathcal{R}e^{-1}+0.03\mu ^{3/2}\left( \frac{\delta }{L}\right) ^{2}\right]
\frac{U^{3}}{L}.
\end{equation*}
\end{theorem}

\begin{proof}
The following uniform in $T$ bounds follow from the energy inequalities and $%
l\leq 0.09\delta $ by differential inequalities as in \cite{KLS21}%
\begin{equation}
\begin{array}{c}
\frac{1}{2}||v(T)||^{2}+\int_{\Omega }k(T)dx\leq C<\infty \text{ ,} \\ 
\frac{1}{T}\int_{0}^{T}\int_{\Omega }\nu |\nabla ^{s}v|^{2}+\nu
_{turb}|\nabla ^{s}v|^{2}+k^{3/2}dxdt\leq C<\infty .%
\end{array}
\label{eq:aPrioriBounds}
\end{equation}%
Time averaging the energy inequality (\ref{eq:EnergyIneq1EqnModel}) and
using the above \'{a} priori bounds and the Cauchy-Schwarz inequality gives%
\begin{equation}
\mathcal{O}\left( \frac{1}{T}\right) +\left\langle \varepsilon _{\text{model}%
}\right\rangle _{T}\leq F\left\langle \frac{1}{|\Omega |}||v||^{2}\right%
\rangle _{T}^{\frac{1}{2}}.  \label{eq:FirstStep}
\end{equation}%
To bound $F$ in terms of flow quantities, take the $L^{2}(\Omega )$ inner
product of the model momentum equation with $f$, integrate by parts and
average over $[0,T]$. This gives%
\begin{gather}
F^{2}=\frac{1}{T}\frac{1}{|\Omega |}(v(T)-v_{0},f)-\left\langle \frac{1}{%
|\Omega |}(vv,\nabla ^{s}f)\right\rangle _{T} \\
+\left\langle \frac{1}{|\Omega |}\int_{\Omega }2\nu \nabla ^{s}v:\nabla
^{s}f+2\nu _{turb}\nabla ^{s}v:\nabla ^{s}fdx\right\rangle _{T}.  \notag
\end{gather}%
The term $\frac{1}{T}\frac{1}{|\Omega |}(v(T)-v_{0},f)$ on the RHS is\ $%
\mathcal{O}(1/T)$. The second term is bounded by the Cauchy-Schwarz
inequality and (\ref{eq:PropertiesL}) by 
\begin{gather*}
\text{ }\left\vert \left\langle \frac{1}{|\Omega |}(vv,\nabla
^{s}f)\right\rangle _{T}\right\vert \leq \left\langle ||\nabla
^{s}f||_{\infty }\frac{1}{|\Omega |}||vv||^{2}\right\rangle _{T} \\
\leq ||\nabla ^{s}f||_{\infty }\left\langle \frac{1}{|\Omega |}||v(\cdot
,t)||^{2}\right\rangle _{T}\leq \frac{F}{L}\left\langle \frac{1}{|\Omega |}%
||v(\cdot ,t)||^{2}\right\rangle _{T}.
\end{gather*}%
The third term is bounded analogously 
\begin{gather*}
\text{ }\left\langle \frac{1}{|\Omega |}\int_{\Omega }2\nu \nabla
^{s}v(x,t):\nabla ^{s}f(x)dx\right\rangle _{T}\leq \left\langle \frac{4\nu
^{2}}{|\Omega |}||\nabla ^{s}v||^{2}\right\rangle _{T}^{\frac{1}{2}%
}{}\left\langle \frac{1}{|\Omega |}||\nabla ^{s}f||^{2}\right\rangle _{T}^{%
\frac{1}{2}}{} \\
\leq \left\langle \frac{2\nu }{|\Omega |}||\nabla ^{s}v||^{2}\right\rangle ^{%
\frac{1}{2}}\frac{\sqrt{2\nu }F}{L}\leq \frac{\beta F}{2U}\left\langle \frac{%
2\nu }{|\Omega |}||\nabla ^{s}v||^{2}\right\rangle _{T}+\frac{1}{\beta }%
\frac{\nu UF}{L^{2}},
\end{gather*}%
for any $0<\beta <1$. The fourth term's estimation is by successive
applications of the space then time Cauchy-Schwarz inequality as follows%
\begin{gather*}
\text{ }\left\vert \left\langle \frac{1}{|\Omega |}\int_{\Omega }2\nu
_{turb}\nabla ^{s}v(x,t):\nabla ^{s}f(x)dx\right\rangle _{T}\right\vert \\
\leq \left\langle \frac{1}{|\Omega |}\int_{\Omega }\left( \sqrt{2\nu _{turb}}%
\right) \left( \sqrt{2\nu _{turb}}|\nabla ^{s}v|\right) |\nabla
^{s}f|dx\right\rangle _{T} \\
\leq ||\nabla ^{s}f||_{\infty }\left\langle \left( \frac{1}{|\Omega |}%
\int_{\Omega }2\nu _{turb}dx\right) ^{\frac{1}{2}}\left( \frac{1}{|\Omega |}%
\int_{\Omega }2\nu _{turb}|\nabla ^{s}v|^{2}dx\right) ^{\frac{1}{2}%
}dx\right\rangle _{T} \\
\leq \frac{F}{L}\left\langle \frac{U}{F}\frac{1}{|\Omega |}\int_{\Omega
}2\nu _{turb}dx\right\rangle _{T}^{\frac{1}{2}}\left\langle \frac{F}{U}\frac{%
1}{|\Omega |}\int_{\Omega }2\nu _{turb}|\nabla ^{s}v|^{2}dx\right\rangle
_{T}^{\frac{1}{2}}.
\end{gather*}%
The arithmetic-geometric mean inequality then implies%
\begin{gather*}
\left\vert \left\langle \frac{1}{|\Omega |}\int_{\Omega }2\nu _{turb}\nabla
^{s}v(x,t):\nabla ^{s}f(x)dx\right\rangle _{T}\right\vert \\
\leq \frac{\beta }{2}\frac{F}{U}\left\langle \frac{1}{|\Omega |}\int_{\Omega
}2\nu _{turb}|\nabla ^{s}v|^{2}dx\right\rangle _{T}+\frac{1}{2\beta }\frac{UF%
}{L^{2}}\left\langle \frac{1}{|\Omega |}\int_{\Omega }2\nu
_{turb}dx\right\rangle _{T}.
\end{gather*}%
Using these four estimates in the bound for $F^{2}$ yields%
\begin{gather*}
F^{2}\leq \mathcal{O}\left( \frac{1}{T}\right) +\frac{F}{L}\left\langle 
\frac{1}{|\Omega |}||v||^{2}\right\rangle _{T}+\frac{1}{2\beta }\frac{UF}{%
L^{2}}\left\langle \frac{1}{|\Omega |}\int_{\Omega }2\nu
_{turb}dx\right\rangle _{T} \\
+\frac{1}{\beta }\frac{\nu UF}{L^{2}}+\frac{\beta F}{2U}\left\langle
\varepsilon _{\text{model}}\right\rangle _{T}.
\end{gather*}%
Thus, we have the estimate%
\begin{gather*}
F\left\langle \frac{1}{|\Omega |}||v||^{2}\right\rangle _{T}^{\frac{1}{2}%
}{}\leq \mathcal{O}\left( \frac{1}{T}\right) +\frac{1}{L}\left\langle \frac{1%
}{|\Omega |}||v||^{2}\right\rangle _{T}^{\frac{3}{2}}{} \\
+\frac{\beta }{2}\frac{\left\langle \frac{1}{|\Omega |}||v||^{2}\right%
\rangle _{T}^{\frac{1}{2}}{}}{U}\left\langle \varepsilon _{\text{model}%
}\right\rangle _{T}+\frac{1}{2\beta }\left\langle \frac{1}{|\Omega |}%
||v||^{2}\right\rangle _{T}^{\frac{1}{2}}{}\frac{2\nu U}{L^{2}} \\
+\frac{1}{2\beta }\left\langle \frac{1}{|\Omega |}||v||^{2}\right\rangle
_{T}^{\frac{1}{2}}{}\frac{U}{L^{2}}\left\langle \frac{1}{|\Omega |}%
\int_{\Omega }2\nu _{turb}dx\right\rangle _{T}.
\end{gather*}%
Inserting this on the RHS of (\ref{eq:FirstStep}) yields%
\begin{gather}
\left\langle \varepsilon _{\text{model}}\right\rangle _{T}\leq \mathcal{O}%
\left( \frac{1}{T}\right) +\frac{1}{L}\left\langle \frac{1}{|\Omega |}%
||v||^{2}\right\rangle _{T}^{\frac{3}{2}}{}  \label{eq:SecondStep} \\
+\frac{\beta }{2}\frac{\left\langle \frac{1}{|\Omega |}||v||^{2}\right%
\rangle _{T}^{\frac{1}{2}}{}}{U}\left\langle \varepsilon _{\text{model}%
}\right\rangle _{T}+\frac{1}{2\beta }\left\langle \frac{1}{|\Omega |}%
||v||^{2}\right\rangle _{T}^{\frac{1}{2}}{}U\frac{2\nu }{L^{2}}  \notag \\
+\frac{1}{2\beta }\left\langle \frac{1}{|\Omega |}||v||^{2}\right\rangle
_{T}^{\frac{1}{2}}{}\frac{U}{L^{2}}\left\langle \frac{1}{|\Omega |}%
\int_{\Omega }2\nu _{turb}dx\right\rangle _{T}.  \notag
\end{gather}%
The last term on the RHS is bounded using H\"{o}lder's inequality as%
\begin{gather*}
\left\langle \frac{1}{|\Omega |}\int_{\Omega }2\nu _{turb}dx\right\rangle
_{T}=\left\langle \frac{1}{|\Omega |}\int_{\Omega }2\mu l\sqrt{k}%
dx\right\rangle _{T}\leq 2\mu 0.09\delta \left\langle \frac{1}{|\Omega |}%
\int_{\Omega }1\cdot \sqrt{k}dx\right\rangle _{T} \\
\leq 2\mu 0.09\delta \frac{1}{T}\int_{0}^{T}\left( \frac{1}{|\Omega |}%
\int_{\Omega }k^{3/2}dx\right) ^{1/3}dt\leq 2\mu 0.09\delta \left\langle 
\frac{1}{|\Omega |}\int_{\Omega }k^{3/2}dx\right\rangle _{T}^{1/3}{}
\end{gather*}%
The second equation of (\ref{eq:EnergyIneq1EqnModel}), the integrated $k-$%
equation, states%
\begin{equation}
\frac{d}{dt}\int_{\Omega }kdx+\int_{\Omega }\frac{1}{l}k^{3/2}dx=\int_{%
\Omega }2\nu _{turb}|\nabla ^{s}v|^{2}dx.
\end{equation}%
Time averaging the above gives%
\begin{equation*}
\mathcal{O}\left( \frac{1}{T}\right) +\frac{1}{T}\int_{0}^{T}\frac{1}{%
|\Omega |}\int_{\Omega }\frac{1}{l}k^{3/2}dxdt=\frac{1}{T}\int_{0}^{T}\frac{1%
}{|\Omega |}\int_{\Omega }2\nu _{turb}|\nabla ^{s}v|^{2}dxdt\text{.}
\end{equation*}%
Thus,%
\begin{equation*}
(0.09\delta )^{-1}\left\langle \int_{\Omega }k^{3/2}dx\right\rangle \leq
\left\langle \int_{\Omega }\frac{1}{l}k\sqrt{k}dx\right\rangle =\left\langle
\int_{\Omega }2\nu _{turb}|\nabla ^{s}v|^{2}dx\right\rangle
\end{equation*}%
and therefore, using (\ref{eq:AnIneq}),%
\begin{eqnarray*}
\left\langle \frac{1}{|\Omega |}\int_{\Omega }2\nu _{turb}dx\right\rangle
&\leq &\mathcal{O}\left( \frac{1}{T}\right) +2\mu 0.09\delta \left\langle 
\frac{1}{|\Omega |}\int_{\Omega }k^{3/2}dx\right\rangle {}^{1/3} \\
&\leq &\mathcal{O}\left( \frac{1}{T}\right) +2\mu (0.09\delta
)^{4/3}\left\langle \frac{1}{|\Omega |}\int_{\Omega }2\nu _{turb}|\nabla
^{s}v|^{2}dx\right\rangle {}^{1/3}
\end{eqnarray*}%
Assembling the above pieces we have%
\begin{gather*}
\left\langle \varepsilon _{\text{model}}\right\rangle _{T}\leq \mathcal{O}%
\left( \frac{1}{T}\right) +\frac{1}{L}\left\langle \frac{1}{|\Omega |}%
||v||^{2}\right\rangle _{T}^{3/2}{} \\
+\frac{\beta }{2}\frac{\left\langle \frac{1}{|\Omega |}||v||^{2}\right%
\rangle _{T}^{1/2}{}}{U}\left\langle \varepsilon _{\text{model}%
}\right\rangle _{T}+\frac{1}{2\beta }\left\langle \frac{1}{|\Omega |}%
||v||^{2}\right\rangle _{T}^{1/2}{}U\frac{2\nu }{L^{2}} \\
\frac{1}{2\beta }\left\langle \frac{1}{|\Omega |}||v||^{2}\right\rangle
_{T}^{1/2}{}\frac{U}{L^{2}}2\mu (0.09\delta )^{4/3}\left\langle \frac{1}{%
|\Omega |}\int_{\Omega }2\nu _{turb}|\nabla ^{s}v|^{2}dx\right\rangle
{}^{1/3}.
\end{gather*}%
Let $T_{j}\rightarrow \infty $, 
\begin{gather*}
\left\langle \varepsilon _{\text{model}}\right\rangle \leq \frac{U^{3}}{L}+%
\frac{\beta }{2}\left\langle \varepsilon _{\text{model}}\right\rangle +\frac{%
1}{2\beta }U^{2}\frac{2\nu }{L^{2}}+ \\
\frac{1}{2\beta }\frac{U^{2}}{L^{2}}2\mu (0.09\delta )^{4/3}\left\langle 
\frac{1}{|\Omega |}\int_{\Omega }2\nu _{turb}|\nabla
^{s}v|^{2}dx\right\rangle {}^{1/3},\text{ } \\
\text{which implies } \\
\left\langle \varepsilon _{\text{model}}\right\rangle \leq \frac{U^{3}}{L}+%
\frac{\beta }{2}\left\langle \varepsilon _{\text{model}}\right\rangle +\frac{%
1}{2\beta }U^{2}\frac{2\nu }{L^{2}}+\left( \frac{1}{2\beta }\frac{U^{2}}{%
L^{2}}2\mu (0.09\delta )^{4/3}\right) \left\langle \varepsilon _{\text{model}%
}\right\rangle {}^{1/3}.
\end{gather*}%
For the last term on the RHS use $ab\leq \frac{2}{3}a^{3/2}+\frac{1}{3}b^{3}$

\begin{eqnarray*}
\left\langle \varepsilon _{\text{model}}\right\rangle &\leq &\frac{U^{3}}{L}+%
\frac{\beta }{2}\left\langle \varepsilon _{\text{model}}\right\rangle +\frac{%
1}{2\beta }U^{2}\frac{2\nu }{L^{2}} \\
&&+\frac{2}{3}\left( \frac{1}{2\beta }\frac{U^{2}}{L^{2}}2\mu (0.09\delta
)^{4/3}\right) ^{3/2}+\frac{1}{3}\left\langle \varepsilon _{\text{model}%
}\right\rangle {}.
\end{eqnarray*}%
Note that $U^{2}\frac{\nu }{L^{2}}=\frac{U^{3}}{L}\frac{\nu }{LU}=\mathcal{R}%
e^{-1}\frac{U^{3}}{L}$. We then have, picking $\beta =2/3,$ collecting like
terms and simplifying, 
\begin{equation*}
\left\langle \varepsilon _{\text{model}}\right\rangle \leq \left[ 3+\frac{9}{%
2}\mathcal{R}e^{-1}+0.03\mu ^{3/2}\left( \frac{\delta }{L}\right) ^{2}\right]
\frac{U^{3}}{L}.
\end{equation*}
\end{proof}

\section{Conclusions and open problems}

The upper bounds for energy dissipation rates of Doering, Constantin and
Foias for the NSE were a breakthrough, connecting turbulence phenomenology
with rigorous mathematical analysis. The recent result of Chow and Pakzad 
\cite{CP20} that expected values of dissipation rates are in some regimes
bounded \textit{below} uniformly in the Reynolds number is a significant
extension of\ this analysis. For turbulence models an upper bound of $%
\mathcal{O}(U^{3}/L)$, where the hidden constant does not blow up as $%
\mathcal{R}e\rightarrow \infty $, directly addresses a question of
computational practice since this estimate precludes a common failure for
the specific model analyzed. Since these are upper bounds, a bound where the
hidden constant blows up as $\mathcal{R}e\rightarrow \infty $, while
suggestive of over dissipation, must be complemented by numerical tests
checking if the upper estimate is sharp.

For $1-$equation or $2-$equation eddy viscosity models and more general flow
problems than shear flows and turbulence in a box, our results suggest
combining an interior cap on the turbulence length scale and a near wall cap
on $\nu _{turb}$ will be effective in many cases. This means that eddy
viscosities $\nu _{turb}$ developed for model accuracy (possibly involving
many calibration constants or using machine learning tools) can be forced to
have a correct energy dissipation balance by essentially $2$ extra lines of
code.

The analysis herein does not include interior shear flows, as when a jet of
fluid enters a large tank, and likely other cases. These are important open
problems. Another common modelling technique is to relax the no slip
condition at walls by a slip with friction / Navier slip law where the
friction is a model calibration coefficient. Analysis of these cases is also
an important open problem.

% \bibliographystyle{plain}
% \bibliography{kean}

\end{document}